\documentclass[conference]{IEEEtran}
\IEEEoverridecommandlockouts

\usepackage{cite}
\usepackage{amsmath,amssymb,amsfonts}
\usepackage{algorithmic}
\usepackage{graphicx}
\usepackage{textcomp}
\usepackage{xcolor}
\usepackage{svg}
\usepackage{booktabs}
\def\BibTeX{{\rm B\kern-.05em{\sc i\kern-.025em b}\kern-.08em
    T\kern-.1667em\lower.7ex\hbox{E}\kern-.125emX}}
\begin{document}

\title{Robust Stuttering Detection via Multi-task and Adversarial Learning}

\author{\IEEEauthorblockN{Shakeel A.~Sheikh$^1$, Md Sahidullah$^1$, Fabrice Hirsch$^2$, Slim Ouni$^1$}
\IEEEauthorblockA{$^1$\textit{Universit\'{e} de Lorraine, CNRS, Inria, LORIA, F-54000, Nancy, France} \\ $^2$\textit{Universit\'{e} Paul-Val\'{e}ry Montpellier, CNRS, Praxiling, Montpellier, France}}
}

\maketitle

\begin{abstract}
By automatic detection and identification of stuttering, speech pathologists can track the progression of disfluencies of persons who stutter (PWS). In this paper, we investigate the impact of multi-task (MTL) and adversarial learning (ADV) to learn robust stutter features. This is the first-ever preliminary study where MTL and ADV have been employed in stuttering identification (SI). We evaluate our system on the \emph{SEP-28k} stuttering dataset consisting of $\approx$ 20 hours of data from 385 podcasts. Our methods show promising results and outperform the baseline in various disfluency classes. We achieve up to 10\%, 6.78\%, and 2\% improvement in repetitions, blocks, and interjections respectively over the baseline. 
\end{abstract}

\vspace{-0.2cm} 
\section{Introduction}
\label{sec:intro}
Speech impairments are speech disorders in which a speaker toils to create the speech sounds required for communication~\cite{guitar2013stuttering}.~These speech impairments can take different forms which include apraxia, cluttering, dysarthria, stuttering, etc~\cite{sheikh2021machine}. Among the speech disorders, stuttering~-~also known by the name stammering/disfluency~-~ has been found to be the most common one~\cite{sheikh2021machine}. Stuttering is a neuro-developmental speech impairment which occurs due to the malfunctioning of sensorimotors, responsible for speech production~\cite{smith2017stuttering}, and usually take the shape of \emph{core behaviours: blocks, repetitions and prolongations}~\cite{guitar2013stuttering}.

\par 
In conventional stuttering detection (SD) paradigm, the speech or recording of PWS is manually analyzed and monitored by speech therapists. However, this method of detecting stuttering is very arduous and time-intensive, and is also prejudiced towards the subjective belief of speech therapist. Besides, automatic speech recognition (ASR) tools also fail to recognize the stuttered speech~\cite{mitra21_interspeech}, that makes it unrealistic to easily access virtual assistants like Alexa, Apple Siri, Cortana, etc. for PWS.~The SD can be utilized to adapt and enhance ASR towards stuttered speech. 
 
\par 
The advancements of deep learning (DL) has shown immense development in various speech domains like speech synthesis~\cite{speechsynthesis}, ASR~\cite{speechrecog}, emotion detection~\cite{speechemotion}, speaker identification~\cite{haqiadversarial}, etc, however SI/SD has received minimal attention from the DL paradigm.~DL can help in detecting the various types of stuttering by exploiting the presence of various acoustic cues in stuttered speech~\cite{sheikh2021machine}.
The recent works have made significant strides in DL based SD~\cite{sheikh2021machine}, however, all the existing systems focus only on single task stuttering learning strategy without focusing on the auxiliary tasks.~Unfortunately, this is not the approach that we human beings process the speech utterances, rather we simultaneously decode the primary task with other meta contents like speaker characteristics, linguistic content, etc. This \emph{multi-task} learning (MTL) have been proven effective in computer vision, ASR, emotion detection, by jointly learning several tasks through a learned shared encoding~\cite{mtlv}.   In this work, we investigate the impact of MTL approach in SD, by jointly learning stuttering and metada information.
\par 
SD systems are further affected by source diversity, be that linguistic content, speaker, gender, accent, or encompassing acoustic conditions~\cite{sheikh2021machine}.~In this study, in addition to MTL, we also investigate an ADV framework with the aim to learn robust stuttering representations.~Among the various source variabilities, we specifically focus on eliminating podcast (meta-data) information with the intention to improve the detection performance of stuttering and its types.~Inspired by unsupervised domain adaptation training~\cite{pmlr-v37-ganin15} and speaker invariant affective learning~\cite{haqiadversarial}, we explore and provide the first ever deeper analysis by using MTL and ADV in the context of SD. 
\par 
Our main contributions are:
\begin{enumerate}
    \item We investigate the effect of applying MTL and ADV in the context of SD. The preliminary results show that the MTL boosts the detection performance of only disfluent classes, and increases the confusion rate for fluent samples. 
    \item We observed that the ADV framework learns robust stutter features, which are stutter discriminate but at the same time are metada invariant  as shown in Fig.~\ref{spkembed}.
    \item To address class imbalance, we use a multi branch training scheme.
\end{enumerate}

\section{Motivation and Related Work}
In this section, we mention the related work and describe briefly the motivation for using MTL and ADV frameworks in the context of SD. 
\vspace{-0.1cm}
\subsection{Motivation}
\label{sec:moti}
\begin{figure}
    \centering
    \includegraphics[scale=0.4]{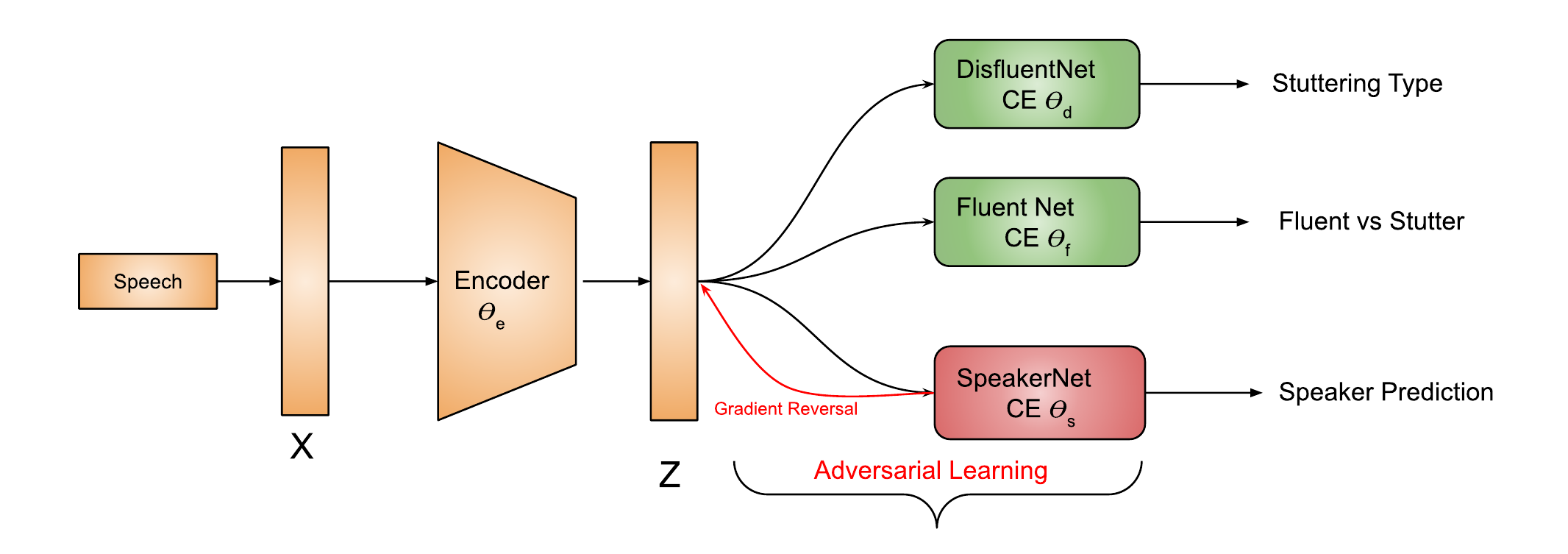} 
    \caption{\small Robust SD using ADV MTL with red curve indicating gradient reversal, CE: Cross entropy, $\theta$'s are the parameters of each module, $X$: MFCC input $\mathcal{R}^{20\times T}$, $Z$: encoder output, $E(X)$.}
    \label{fig:advmtl} \vspace{-0.4cm}
\end{figure}

Multi-task learning (MTL) is basically an inductive transfer scheme which simultaneously learn multiple tasks by optimizing several objectives using a shared hidden representation~\cite{caruana1997multitask, mtlv}. It provides an effective way of improving generalization. MTL aims to exploit the integrated knowledge across multiple domains with the intention of improving the performance of primary task~\cite{crawshaw2020multi}. MTL has been successfully applied in several tasks in speech domain such as emotion detection, speaker recognition, ASR~\cite{mtlv}, etc. 

\par 
A contrasting approach is to presume that robust acoustic representation of stuttering should be invariant to secondary tasks, in particular speaker recognition, as stuttering speech task should not depend on a particular set of speakers. The way to achieve such robust invariances is ADV learning~\cite{pmlr-v37-ganin15}. The ADV framework learns invariances to metada characteristics in a min-max fashion that the main branch is trained to maximize the metadata (podcast) classification loss and the speaker branch is trained to minimize its classification loss. Robust representation of speech signals have been studied via ADV training in various speech domains such as emotion detection~\cite{haqiadversarial} and ASR~\cite{asradv}. However, there is no prior study on the application of MTL and ADV frameworks in the domain of acoustic SD. 
\vspace{-0.1cm}

\subsection{Related Work}
\vspace{-0.1cm}
\label{sec:rework}
Most of the work related to SD employ mel-frequency cepstral coefficients (MFCCs) or spectrograms as feature extraction methods~\cite{sheikh2021machine}. T. Kourkounakis \emph{et al}~\cite{tedd} used ResNet and BiLSTM to solve the problem of SD. The latter was formulated as a multiple binary classification problem, where they trained separate model for each type of stuttering.~In another study, they proposed FluentNet for end-to-end identification of stuttered speech~\cite{fluentnet}. The networks were trained using the spectrograms, which were the sole input features to the model. The proposed architecture showed promising results on a small subset of speakers (25) of UCLASS dataset, but did not perform well on a large set of speakers~\cite{stutternet}. In addition, they have mentioned it end-to-end, but they are still using the hand crafted spectrogram features.  Recently Lee \emph{et al}~\cite{sep28k} introduced ConvLSTM model for detecting  various types of stuttering: blocks, interjections, word repetitions, and sound repetitions. The architecture takes 40 MFCCs with pitch and articulatory features as an input, and was trained with a batch size of 256 using cross entropy loss function. They also introduced the new stuttering SEP-28k dataset with metadata information, which we exploit in this study.
Melanie~\emph{et al.}~\cite{melanie} recently introduced the phoneme based BiLSTM for disfluency detection. Even though the model shows promising results but they have mixed all the datasets together for their case study. We recently introduced a single branch multi-class~\emph{StutterNet}, a time-delay neural network based stuttering classifier, capable of identifying fluent segment and core behaviours~\cite{stutternet}, and, it shows encouraging results compared to state-of-the-art on a large set of speakers. The \emph{StutterNet} was trained with MFCCs and cross entropy on a set of 104 speakers from UCLASS dataset. A detailed summary of various SD methods can be found in the review paper by Shakeel \emph{et al}~\cite{sheikh2021machine}. 
\par 
This study is an extension of our previous work~\emph{StutterNet}~\cite{stutternet}.
Due to the diversity and uniqueness of stuttering, it continues to be the most demanding and challenging to detect, due to its inherent nature of huge class imbalance across different types of disfluencies~\cite{sheikh2021machine}. Training a model on a such type of dataset will bias the majority class. To this end, we employ a MB \emph{StutterNet} (baseline), where, we introduce the pseudo temporary labelling of all disfluent classes as one class. The MB \emph{StutterNet} is a two branched network, with one branch for differentiating between fluent vs disfluent and the other branch differentiates among various disfluent types. In addition, we investigate the effectiveness of MTL in SI/SD domain, by jointly training the primary task i.e, stuttering classification and auxiliary task i.e, podcast classification. The MTL has shown promising results in other domains of speech such as emotion detection, speaker recognition, ASR, etc.~\cite{mtlv}. However, no attention has been given regarding the impact of adapting MTL framework in SD. This paper proposes a SI/SD method that utilizes TDNN based MTL and ADV with the intention to improve generalization performance of primary task (SD) by learning jointly the metada information.

\vspace{-0.06cm}
\section{Proposed Framework}
\vspace{-0.1cm}
This section describes the framework for MTL and ADV schemes that we have employed in this case study.
\label{sec:proposed}
\subsection{Multi-Task Learning}
Within the MTL framework, given a training set of $\mathcal{D} = (X_i, s_i, d_i)_{i=0}^{n}$, where each sample is comprised of three components: a sequence of $\mathcal{R}^{20\times T}$ acoustic MFCC features, a speaker label $s_i$, pseudo fluent label $f_i$ (for \emph{FluentNet}, all different stuttering classes are treated as a single disfluent
class by introducing a temporary pseudo-labelling scheme to solve the class imbalance problem) and a stutter label $d_i$, we consider a shared encoder $E$ having parameters $\theta_e$ and three classifier branches $S$ with parameters $\theta_s$, $F$ with parameters $\theta_f$ and $D$ with parameters $\theta_d$ for podcast, fluent and stuttering sub type detections respectively. Our goal is to minimize both stutter loss $\mathcal{L_\text{stutter}(\theta_\text{e}, \theta_\text{d})}$ and speaker (podcast here) loss 
$\mathcal{L_\text{speaker}(\theta_\text{e}, \theta_\text{s})}$.~Therefore, the total objective function in MTL can be represented as:\vspace{-0.4cm}

\begin{equation}
\small
\label{eq:lossmtl}
\begin{split}
\small
   \mathcal{L(\theta_\text{e}, \theta_\text{f}, \theta_\text{d}, \theta_\text{s})}  = &(1 - \lambda) *\mathcal{L_\text{stutter}(\theta_\text{e}, \theta_\text{f}, \theta_\text{d})} + \lambda *\mathcal{L_\text{speaker}(\theta_\text{e}, \theta_\text{s})} \\
    \mathcal{L_\text{stutter}(\theta_\text{e}, \theta_\text{f}, \theta_\text{d})} =& 
  \mathcal{L_\text{fluent}(\theta_\text{e}, \theta_\text{f})} + \mathcal{L_\text{disfluent}(\theta_\text{e}, \theta_\text{d})}
  \end{split}
\end{equation}
where, $\lambda$ $\in$ (0, 1), is a weighting parameter between the losses tuned manually. 
\subsection{Adversarial Learning}
In applying ADV framework to stuttering domain, we aim to learn acoustic representations which are robust and metada invariant, but at the same time is stuttering discriminative. For this purpose, similar to MTL, we consider an encoder $E$ having parameters $\theta_e$, which maps the $T$ input acoustic features
to representation $Z$ by $Z_i = E(X_i)$. This $Z_i$ representation is then fed to the  \emph{DisfluentNet} (denoted by D) and \emph{FluentNet} decoders (denoted by F) to output the  stuttering and fluent class probabilities $p(d_i/Z_i;\theta_e, \theta_d)$ and $p(f_i/Z_i;\theta_e, \theta_f)$ respectively. In addition, this $Z_i$ is fed to another branch speaker decoder (denoted by S), to output metadata (podcast) probabilities $p(s_i/Z_i;\theta_e, \theta_s)$. To learn metadata-invariant representations, our goal is to minimize stuttering classification loss, $\mathcal{L_\text{stutter}(\theta_\text{e}, \theta_\text{d})}$ and to optimize $\theta_e$ to maximize the speaker classification loss, $\mathcal{L_\text{speaker}(\theta_\text{e}, \theta_\text{s})}$  and, at the same time minimize the speaker classification loss by optimizing $\theta_s$. The total objective loss becomes:  
\begin{equation}
 \small
\label{eq:lossadv}
\begin{split}
\small
   \mathcal{L(\theta_\text{e}, \theta_\text{f}, \theta_\text{d}, \theta_\text{s})}  = \mathcal{L_\text{stutter}(\theta_\text{e}, \theta_\text{f}, \theta_\text{d})} - \lambda *\mathcal{L_\text{speaker}(\theta_\text{e}, \theta_\text{s})} \\
       \mathcal{L_\text{speaker}} = -\sum\limits_{(X_i, f_i, d_i,s_i) \in \mathcal{D}}\log(p(s_i/Z_i)) \\
       \mathcal{L_\text{stutter}} = -\sum\limits_{(X_i, f_i, d_i,s_i) \in \mathcal{D}}\log(p(d_i/Z_i)) + \log(p(f_i/Z_i)) 
   \end{split}
\end{equation}
where, $\lambda$ $\in$ (0, 1), is a weighting trade-off hyper-parameter, and $s_i$, $f_i$ and $d_i$ are the labels. We optimize the parameters using Adam by applying the gradient reversal layer~\cite{pmlr-v37-ganin15} in speaker branch during backpropagation as shown by red curve in Fig.~\ref{fig:advmtl}. 

 \vspace{-0.2cm}
 \subsection{Model Architecture} 
 \vspace{-0.05cm}
 
 Our SD models are based on time delay neural networks, which has been proven effective in various speech domains~\cite{stutternet}. The model is fed with mean-normalized 20-dimensional MFCC input features extracted on a 20~ms sliding window with a hop length of 10~ms. The proposed ADV model is based on MTL framework with four components including the encoder $E$, which generates representations, the \emph{FluentNet} $F$ classifier branch, the stutter classifier module named as \emph{DisfluentNet} $D$, and the speaker classifier $S$ as illustrated in Fig.~\ref{fig:advmtl}. The encoder module is composed of five time delay layers with the first three focusing only on small contextual frames of $[t-2, t+2]$, $\{t-2,t,t+2\}$, and $\{t-3,t,t+3\}$ respectively. The remaining two focuses on the $\{t\}$ contextual frames. The last layer of encoder is statistical pooling layer (SPL) with mean and standard deviation.~The fixed dimension representation from SPL, which is the output of the $E$, is further passed to the three sub-classifiers: \emph{DisfluentNet}, \emph{FluentNet} and \emph{SpeakerNet}. These three sub modules are comprised of three fully connected layers and a softmax layer. A dropout of 0.2 is applied after first two fully connected layers in each branch.~Except SPL, each layer is followed by ReLU non linear activation function and a 1D batch normalization. To address the class imbalance problem, we use two branches \emph{FluentNet} and \emph{DisfluentNet} for SD. During inference time, speaker branch $S_\theta$ is discarded, and \emph{FluentNet} output is considered, if it predicts fluent, otherwise \emph{DisfluentNet} predictions are considered.
 \vspace{-0.15cm}

 \vspace{-0.1cm} 
\section{ Experimental Setup}
 \label{sec:expt}
 \vspace{-0.1cm} 

\textbf{Dataset:} 
For our experimental studies, we have used the SEP-28k stuttering dataset, curated recently by Apple from a set of 385 podcasts across 8 shows~\cite{sep28k}. From each podcast episode, 40-250 three second segments were extracted, resulting in an overall total samples of 28,177.~The SEP-28k dataset is having two different types of annotations: stuttering and non-stuttering. The stuttering annotation consists of \emph{core behaviours}, which include repetitions, prolongations and blocks. In addition to \emph{core behaviour} labels, it also contains annotation for interjections and fluent segments.~The non stuttering annotation is comprised of poor audio quality, music, no speech, unsure, unintelligible and natural pauses. In this experimental study, we used and focused mainly on the stuttering annotations.~Out of 28,177 samples, after leaving out non stuttering samples, we used only 23573 speech segments for our case study which constitutes of 19.65 hours of data. The distribution of stuttering data is highly imbalanced, among which the majority of the samples (12419) are fluent segments constitutes 10.35 hours, 3995 are interjections constitutes 3.34 hours, 3286 are repetitions constitutes 2.74 hours, 2103 are blocks constitutes 1.75 hours, and 1770 are prolongations constitutes 1.48 hours of data. For our baseline, we randomly selected 80\% of podcasts for training, 10\% for validation and remaining 10\% for test set which comprises of 309, 37 and 39 podcasts respectively. For MTL and ADV, we mix the train and validation sets from baseline, and then randomly selected 90\% samples for train set and 10\%  samples for validation set from each podcast to have at-least each disfluency from each podcast in both the sets, and the test set remains same in all the cases. In this study, we use podcast ids as meta deta information to learn robust stutter representations.

\textbf{Evaluation Metrics:} To test and evaluate the model performance, we use the widely and standard metrics that are used in stuttering and other speech tasks~\cite{sheikh2021machine} which include recall, precision, F1-score and accuracy. The models are evaluated and compared to our previous work \emph{StutterNet}~\cite{stutternet}. The results reported are the average of 10 experiments with $10$- fold cross validation technique.~For ADV, we first train the \emph{SpeakerNet} for 25 epochs by freezing the other two branches, then we freeze the \emph{SpeakerNet} and train the \emph{FluentNet} and \emph{DisfluentNet} for 25 epochs. From epoch 50-75, we jointly train all the branches by applying gradient reversal layer in the speaker branch during backpropogation, and after epoch 75, we freeze all the weights and fine tune the \emph{FluentNet} and \emph{DisfluentNet} branches for the recovery phase until the stopping criteria is achieved.

\textbf{Implementation:} We have used PyTorch library for our implementation purposes. We trained the models using Adam optimizer, cross entropy loss function with a learning rate of $10^{-2}$. The training was stopped with an early stopping criteria having a patience of 7 on validation loss. 

\begin{table}[h]
    \centering
    \caption{\small Stuttering detection results (B: Block, F: Fluent, R: Repetition, P: Prolongation, I: Interjection, MB: Multi branch, BL: Baseline)}
    \scalebox{0.8}{\begin{tabular}{*{7}{c}}
    \toprule
        \multicolumn{1}{c}{}& \multicolumn{1}{c}{}&\multicolumn{5}{c}{Precision}  \\
    \midrule
   Model &Method&R&P&B&I&F\\
        \midrule
        \emph{StutterNet}~\cite{stutternet}&SB&0.22&0.28&0.02&0.50&0.88\\
        \emph{MB StutterNet}&BL &0.35	&0.36&	0.23&	0.58&	0.67\\
         \emph{MB StutterNet}&MTL& 0.32	&0.32&	0.10&	0.59&	0.72\\
           \emph{MB StutterNet}&ADV& 0.27&	0.33&	0.08&	0.56&	0.77 \\
           \midrule
           &&&&Recall.&&\\
           \midrule
           \emph{StutterNet}~\cite{stutternet}&SB&0.42&0.42&0.25&0.69&0.62\\
           \emph{MB StutterNet}&BL &	0.29&	0.38&	0.10	&0.58&	0.74\\
         \emph{MB StutterNet}&MTL&0.34&	0.36&	0.21&	0.54&	0.67\\
           \emph{MB StutterNet}&ADV&	0.35&	0.37&	0.20	&0.58&	0.66 \\
           \midrule
           &&&&F1 Score.&&\\
           \midrule
           \emph{StutterNet}~\cite{stutternet}&SB&0.29 &0.33 &0.04 &0.57 &0.73\\
           \emph{MB StutterNet}&BL &	0.31&	0.36&	0.12&	0.57&	0.70\\
         \emph{MB StutterNet}&MTL&	0.32&	0.33&	0.13&	0.56&	0.69\\
           \emph{MB StutterNet}&ADV& 	0.3	&0.34&	0.12&	0.57&	0.71 \\
           \bottomrule
    \end{tabular}}
        \label{tab:precision}
 \end{table}
 
 \begin{table}[]
 \vspace{-0.52cm} 
    \centering
    \caption{\small  Stuttering detection results (SA: Stutter accuracy, TA: Total accuracy, SB: Single branch)}\vspace{-0.25cm}
     \scalebox{0.8}{\begin{tabular}{*{9}{c}}
   
    \toprule
    Model&&R&P&B&I&SA&F&TA\\
        \midrule
        \emph{StutterNet}~\cite{stutternet} &SB&21.99&27.78&1.98&49.99&29.93&88.18&60.33	 \\
        \emph{MB StutterNet} &BL&	28.70&	37.89	&9.58&	57.65&	37.72&	74.43&	57.04  \\
        \emph{MB StutterNet} &MTL&	31.59&	31.62	&10.23	&58.92&	38.32&	72.14&	56.09\\
        \emph{MB StutterNet} &ADV& 27.24&	32.89&	8.33&	56.36&	35.96&	77.10&	57.51 	\\
        \bottomrule
            \end{tabular}}
    \label{tab:Accuracylf}  \vspace{-0.5cm}
 \end{table}

\begin{figure}[h]
\vspace{-0.4cm}
\hspace{-0.5cm}
\begin{minipage}[t]{0.5\linewidth}
    \centering
    \includegraphics[width=1\textwidth]{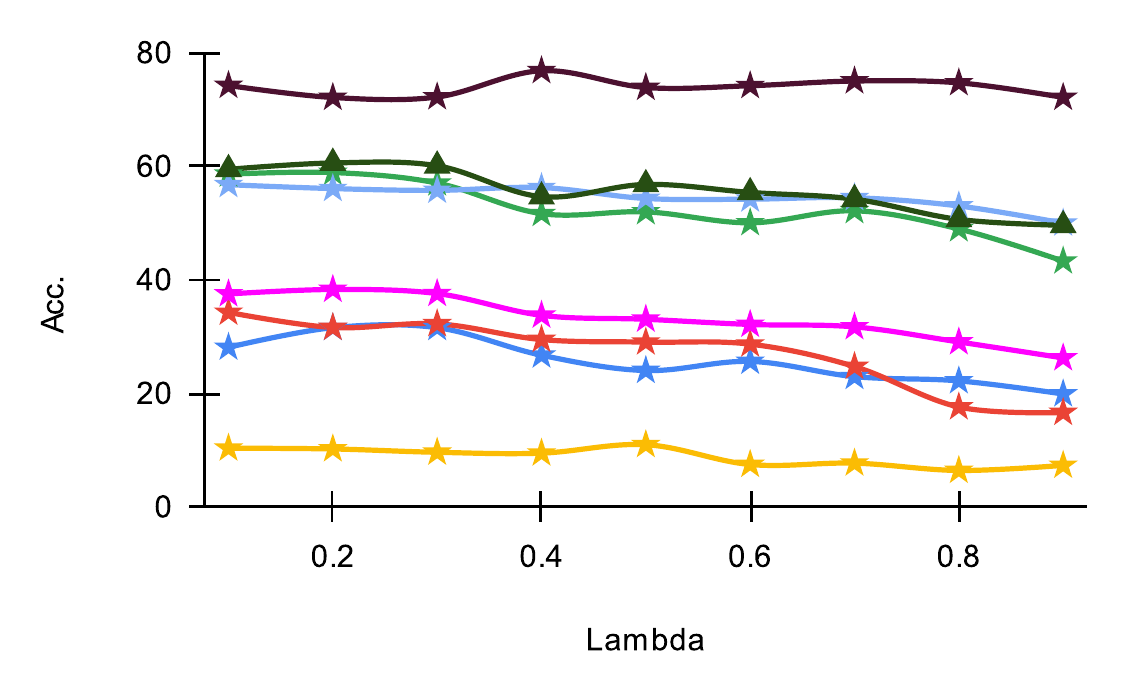} 
\end{minipage}
\hspace{0.2cm}
\begin{minipage}[t]{0.5\linewidth} 
    \centering
    \includegraphics[width=1\textwidth]{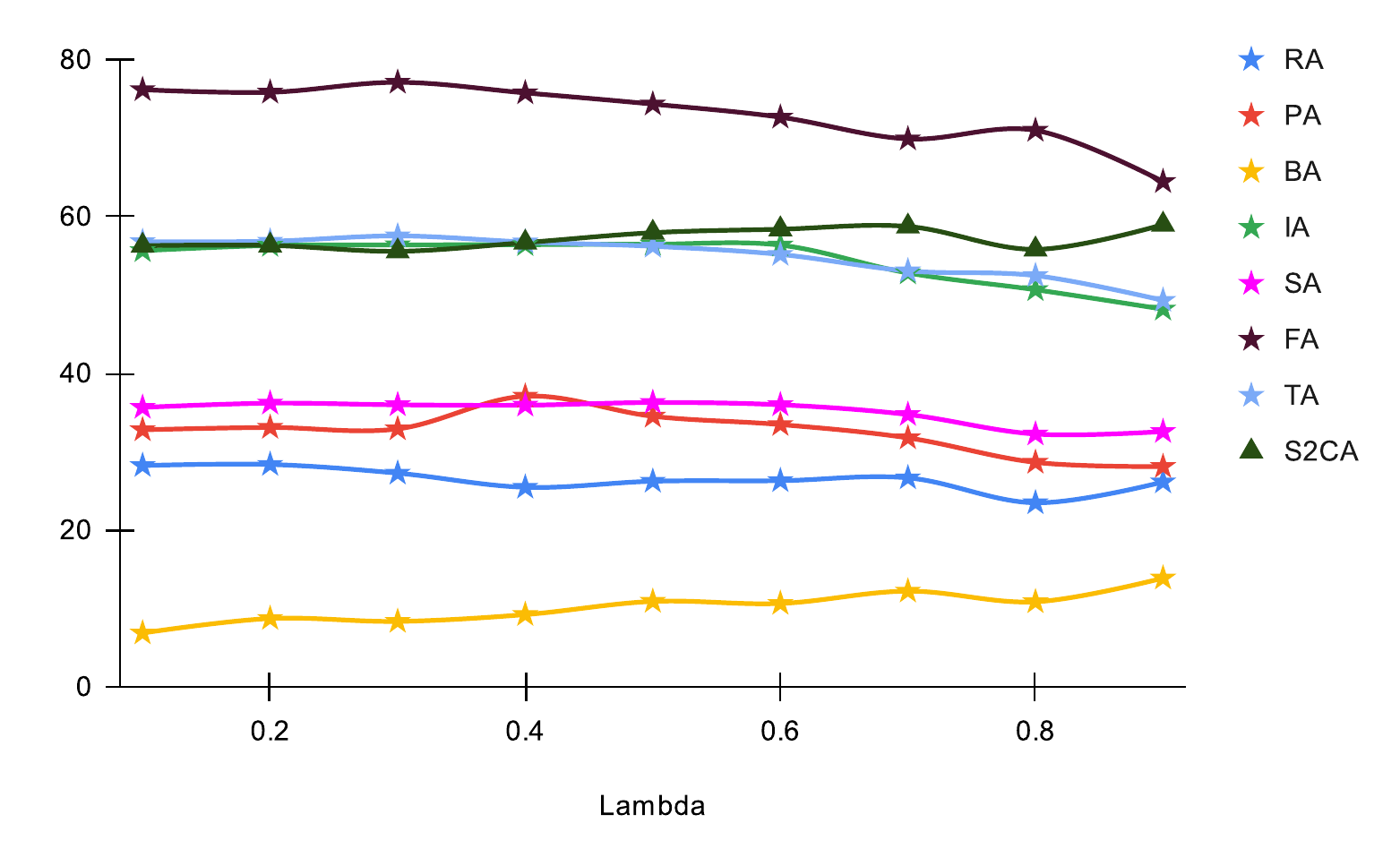}
\end{minipage}   

\vspace{-0.65cm}
\caption{\small Effect of lambda in MTL~(left) and ADV~(right) in SD.} \vspace{-0.3cm}
\label{lambdaeffect}
\end{figure} 
\begin{figure}[h]
\vspace{-0.2cm}
\begin{minipage}[t]{1.0\linewidth}
      \centering
    \includegraphics[width=1\textwidth]{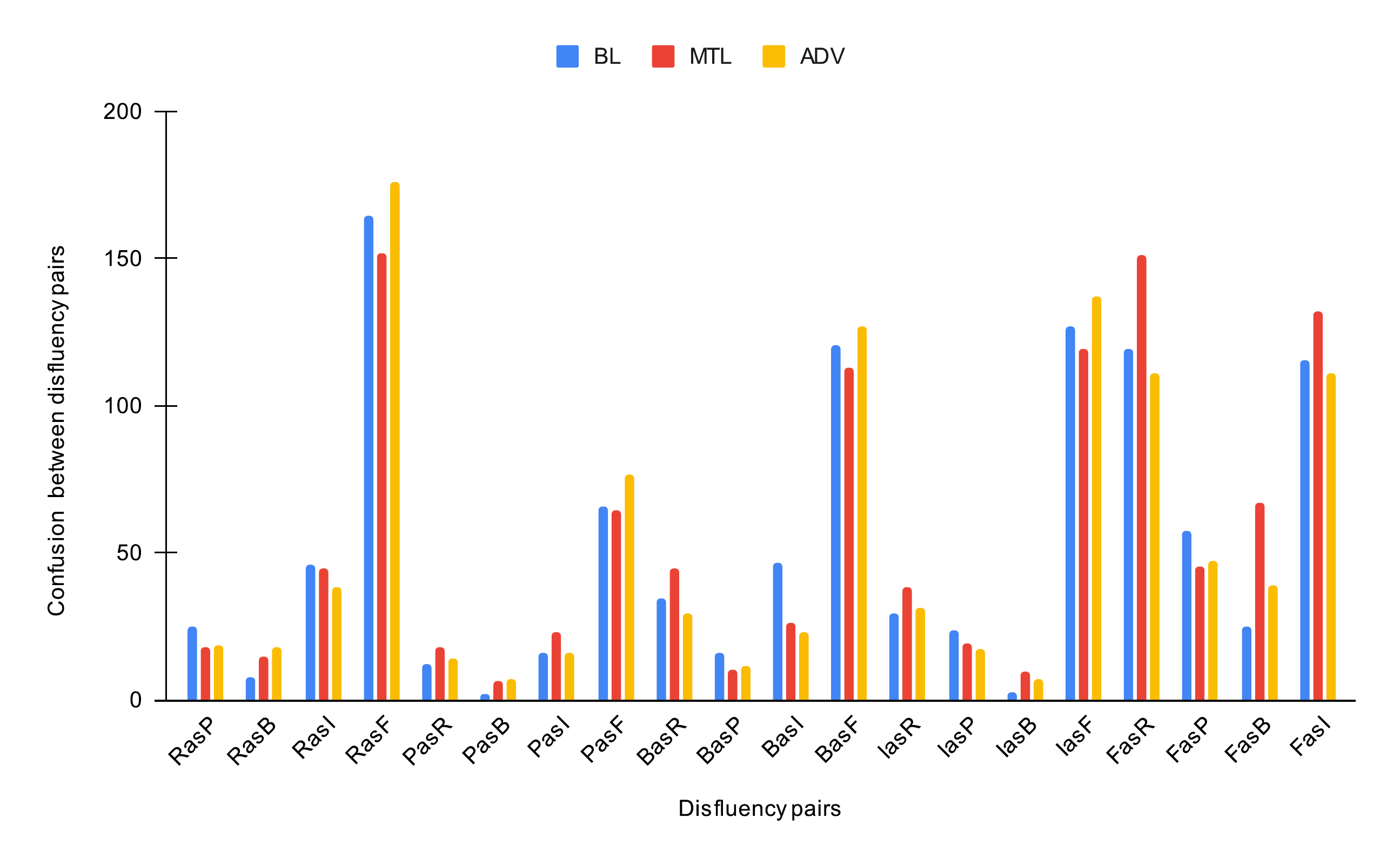}
  \vspace{-0.9cm}\caption{\small Confusion in disfluency pairs in BL, MTL and ADV. XasY refers class X identified as class Y. For better visualization, we did not show in percent.}
\label{confrate} 
\end{minipage}

\vspace{-0.5cm}
\end{figure}

\vspace{-0.13cm}
\section{Results and Discussion}
\label{sec:res}
\vspace{-0.10cm}
\begin{figure*}[h]
 \vspace{-0.5cm}
 \begin{minipage}[t]{0.5\linewidth}
    \centering
    \includegraphics[width=0.9\textwidth]{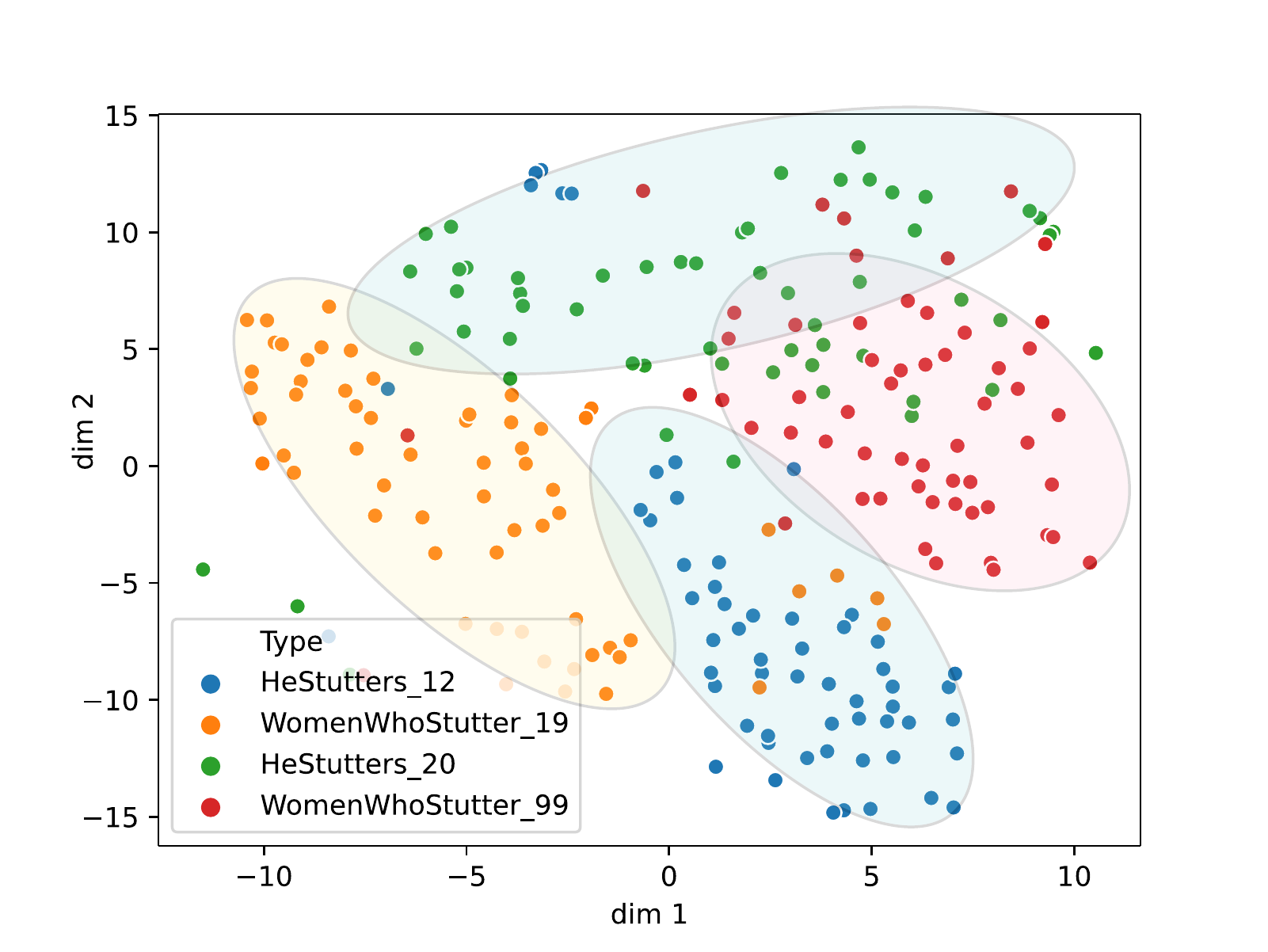}  
   
\end{minipage}
\begin{minipage}[t]{0.5\linewidth} 
    \centering
    \includegraphics[width=0.9\textwidth]{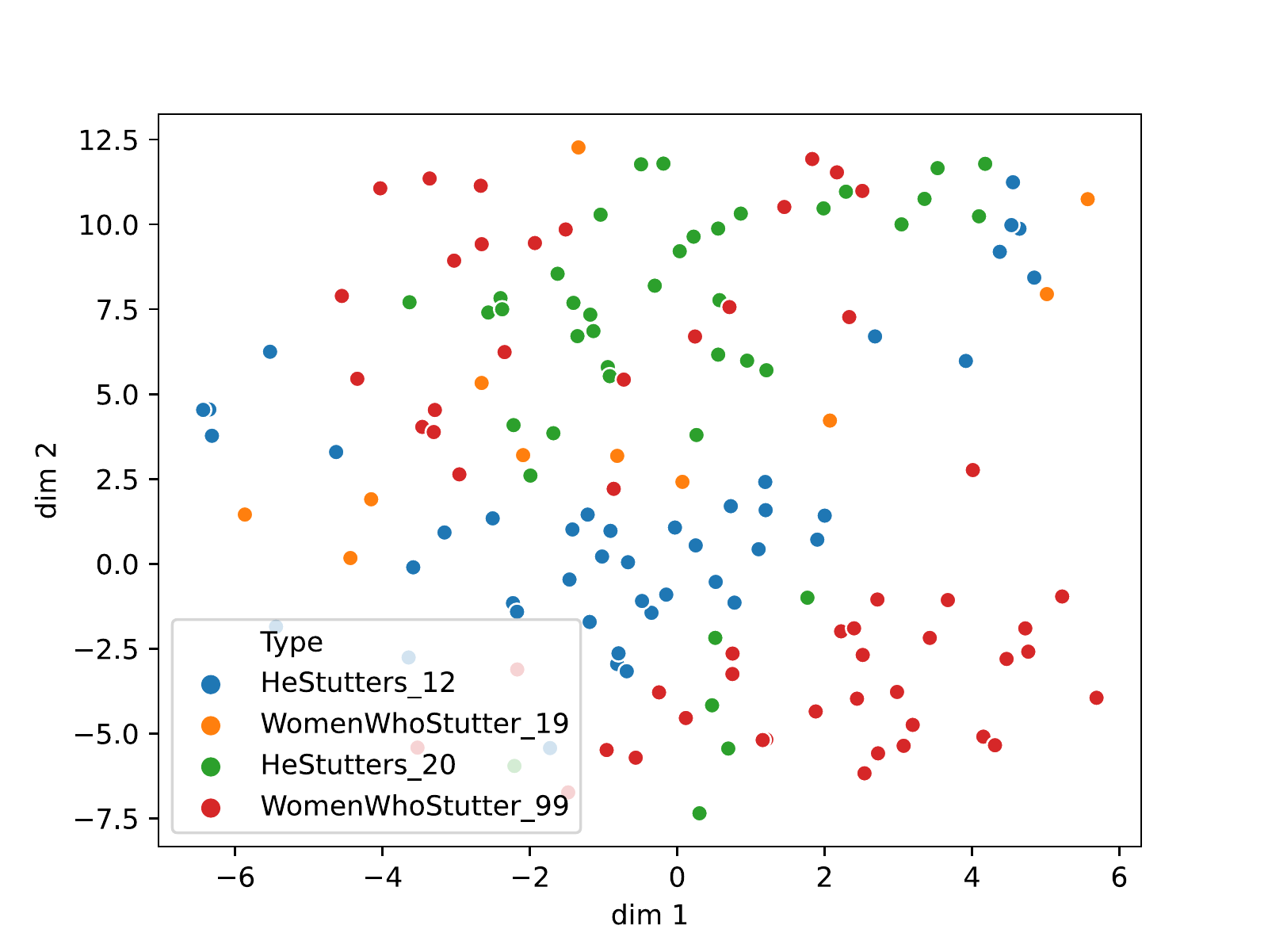}
\end{minipage} 
\vspace{-0.7cm}
\caption{ \small t-SNE embeddings from the shared encoder MTL (L) and ADV (R). We selected four podcasts for visualisation purposes.}
\vspace{-0.5cm}
\label{spkembed}
\end{figure*} 
\textbf{Baseline:} Tables~\ref{tab:precision} and~\ref{tab:Accuracylf} show the results of our baseline StutterNet along with the proposed methods. In comparison with the single branch (SB), multibranch (MB) outperforms the single branch in all the disfluency class predictions with an overall improvement of 26\%. However, the SB shows better performance on fluent class because of the class imbalance issue. After addressing the class imbalance issue via two branch (\emph{FluentNet} and \emph{DisfluentNet}) training scheme, we found that the blocks and repetitions are still being confused with fluent speech as shown in Fig.~\ref{confrate}. This is intuitive because the repetitions are fluent speech with the same word being repeated more than once. Similarly, the blocks are fluent speech with some small initial pause followed by fluent speech. From the Table~\ref{tab:Accuracylf}, we also observe that compared to the single task learning approach, our model based on simple MTL approach helps to boost the detection performance for stuttering types but degrades for fluent and prolongation segments, as the confusion rate of fluent and prolongation samples increases with  other disfluency pairs in the MTL framework as shown by the red bar plots in Fig.~\ref{confrate}. 
The MTL-based model increases the detection performance of R, B and I's by a relative margin of 10\%, 6.78\%, and 2\% respectively, which results in an overall improvement by 1.6\% in the stuttering classes. More interestingly, we found that the ADV improves the detection performance of fluent classes by 3.59\% as compared to the baseline. It can be clearly seen from the Fig.~\ref{confrate}, that there is a notable drop in the confusion rate of fluent category with other classes, but at the same time, the fluent false positive rate also increases. By adversarial training scheme, the model learns podcast-invariant representations which improves the performance of the fluent class.

\par 
\textbf{Effect of lambda:}
Figure.~\ref{lambdaeffect}~(left)\footnote{Initial is disf. class, A: Acc., S2CA: Stutter two class Acc. of \emph{FluentNet}} shows the result of MTL when the value of $\lambda$ hyper-parameter in the loss eq.~(\ref{eq:lossmtl}) is varied from 0.1 to 0.9, and it includes the accuracy of all the classes. The results for all classes almost show a similar trend except for fluent class. The $\lambda=0.4$ shows the best accuracy for the fluent class. As the $\lambda$ value increases, the detection performance of all disfluent classes decrease as well as in fluent class, which is expected because the overall loss function assigns more weight to the podcast identification branch. In addition, we also tried to divide $\lambda$ by 10 (starting with 1) at each epoch, and we found that it increases the detection performance of blocks slightly and repetitions considerably, with slight reduction in other classes. Figure.~\ref{lambdaeffect}~(right) shows the result of ADV training, with varying $\lambda$. The block class follows a trend that by increasing the $\lambda$, its detection performance also improves. For other classes, as the value of $\lambda$ is increased from $0.1\to 0.9$, the identification performance decreases. Following the work in~\cite{grl}, we also tried varying $\lambda$ by $2/(1+e^{p_i}) -1$, where $p_i$ is the scaled version of epoch. However, it degrades performance. 
\par 

In addition, the two class accuracy of disfluent samples in the \emph{FluentNet} decreases as we increase the value of $\lambda$ in the MTL framework. The $\lambda$ acts as a control parameter for the podcast information to flow through the network. As $\lambda$ increases, it is allowing more and more podcast-related information to flow through the network and thus decreasing the detection performance of stutter samples in the \emph{FluentNet}. To this contrary, the podcast identification branch in the ADV framework is trying to play a good adversary and tries to remove more irrelevant information as $\lambda$ value increases. This, in turn, helps in boosting the detection performance of stutter samples in the \emph{FluentNet} branch. We can conclude that the podcast information is entangled with the stuttering characteristics, and by reducing podcast-related information, the model is able to improve the discrimination between the fluent and disfluent classes, but it fails to generalize within the disfluent classes.
\par 
\textbf{Visualization:} Figure.~\ref{spkembed} shows the embeddings computed from the statistics pooling layer. In MTL scheme, it is evident from the well formed podcast clusters that the model is trying to learn podcast dependent stuttering information. On the other hand, for the adversarial setting, the clusters are not visible and the model is trying to learn this meta-data invariant robust stutter features.

\section{Conclusion and Future Work}
\label{sec:conc}
Due to the diversity and uniqueness of stuttering, it continues to be the most demanding and challenging to detect due to its inherent nature of huge class imbalance across different types of disfluencies. Training a model on such a type of dataset will bias the majority class. To this end, we employ an MB \emph{StutterNet} (BL), where, we introduce the pseudo temporary labeling of all disfluent classes as one class. The MB \emph{StutterNet} is a two-branch network, with one branch to differentiate fluents from disfluent and the other branch to differentiate different types of disfluent types if the prediction in the \emph{FluentNet} is non-fluent. 
On top of the baseline, we investigate the MTL and ADV framework by adding one more speaker branch in the context of SD. From the results, we found that the MTL framework improves the detection performance of blocks, repetitions, and interjections. We also found that the confusion rate increases for the fluent and prolongation classes.  With the ADV scheme, we found that the fluent detection performance increases with a relative improvement of 3.59\%. In addition, we also found that by decreasing the $\lambda$ hyper-parameter in both the frameworks, the detection rate of disfluent classes increases. There might be issues with the meta-information of the dataset, as there is a possibility of having the same speakers across multiple podcasts. 
\par 
Future works may include to study and investigate the impact of additional meta-data like language, accent, acoustic conditions in the context of SD.

\footnotesize
\vspace{-0.1cm}
\section*{Acknowledgment}
\vspace{-0.1cm}
This  work  was  made  with  the  support  of  the  French  National  Research Agency, in the framework of the project ANR BENEPHIDIRE (18-CE36-0008-03). Experiments  presented  in  this  paper  were  carried  out  using  the  Grid’5000 testbed, supported by a scientific interest group hosted by Inria and including CNRS,  RENATER  and  several  universities  as  well  as  other  organizations(see  https://www.grid5000.fr) and  using the EXPLOR  centre, hosted by the University of Lorraine.

{\footnotesize \bibliographystyle{IEEEtran}
 \bibliography{refs.bib}  }

\end{document}